\def\beq{\begin{equation}}
\def\eeq{\end{equation}}
\def\beqa{\begin{eqnarray}}
\def\eeqa{\end{eqnarray}}

\def\iv{$I$-$V$}

\documentclass[superscriptaddress,prb,twocolumn,amsmath,amssymb,showpacs]{revtex4}
\usepackage[dvipdfm]{graphicx} 
\usepackage{color,ulem}

\begin{document}

\title{Possible method to observe breathing mode of magnetic domain wall in Josephson junction}

\author{Michiyasu Mori}
\affiliation{Advanced Science Research Center, Japan Atomic Energy Agency, Tokai, Ibaraki 319-1195, Japan}
\affiliation{CREST, Japan Science and Technology Agency,Kawaguchi 433-0012, Japan}

\author{Wataru Koshibae}
\affiliation{RIKEN Center for Emergent Matter Science (CEMS), 2-1 Hirosawa, Wako,
Saitama 351-0198, JAPAN} 
\affiliation{CREST, Japan Science and Technology Agency,Kawaguchi 433-0012, Japan}

\author{Sin-ichi Hikino}
\affiliation{Computational Condensed Matter Physics Laboratory, RIKEN, Wako, Saitama 351-0198, Japan}
\affiliation{CREST, Japan Science and Technology Agency,Kawaguchi 433-0012, Japan}

\author{Sadamichi Maekawa}
\affiliation{Advanced Science Research Center, Japan Atomic Energy Agency, Tokai, Ibaraki 319-1195, Japan}
\affiliation{CREST, Japan Science and Technology Agency,Kawaguchi 433-0012, Japan}

\begin{abstract}
A magnetic domain wall (DW) behaves as a massive particle with elasticity. 
Sliding and oscillation of the DW have been observed experimentally, whereas vibration of a width in the DW, "{\it breathing mode}", has not been measured so far.  
We theoretically propose how to observe the breathing mode by the Josephson junction having a ferromagnetic layer between superconducting electrodes. 
The current-voltage ($I$-$V$) curve is calculated by an equivalent circuit of the resistively shunted junction model. 
The breathing mode is identified by stepwise structures in the $I$-$V$ curve, which appear at the voltages $V=n(\hbar/2e)\omega$ with the fundamental constant $\hbar/e$, integer number $n$, and the frequency of the breathing mode $\omega$. 
\end{abstract}

\date{May 1, 2014}

\maketitle

\section{Introduction}
A magnetic domain wall (DW) is a solitonic object which  
connects two stable configurations of magnetizations. 
Many studies are devoted to control the DW due to many advantages 
from an application viewpoint 
such as enhanced operation speed, low power consumption, and high density integration 
of electronic devices~\cite{Chikazumi,Maekawa,Doring,saitoh, Yamano,Yamagu,Thomas,hayashi,boone,bisig,Thia,he07,Tata,ieda}.
Experimentally, it is shown that the DW behaves as a particle with finite mass~\cite{Doring,saitoh} and is driven by an electric current~\cite{Yamano,Yamagu}. 
Oscillation of the DW confined in a potential is studied experimentally~\cite{saitoh,Thomas,hayashi,boone,bisig} and theoretically~\cite{Thia,he07,Tata,ieda,Hiki12}. 
In addition to the oscillation, 
a vibration of a width of the DW called $\lq\lq${\it breathing mode}$"$ is expected, 
since this solitonic object is composed of many electron spins. 
In the previous theoretical studies~\cite{Dant,Jung,Mats09,Mats12}, the breathing mode induced by the electric current has been discussed.
A current-driven breathing mode is expected to be a microwave source 
of 
wireless telecommunication devices~\cite{Mats09,Mats12}. 
Experimentally, however, the method to detect and measure the breathing mode of DW has not been established so far.

Superconductors (SCs) provide many types of devices sensitive to an external magnetic field 
and are used for precise measurement of voltage~\cite{Baro,Tink}. 
Superconducting quantum interference devices (SQUIDs) presently offer the highest sensitivity to a magnetic field.  
The Josephson junction under irradiation of microwave provides a precise value of voltage adopted to the voltage standard~\cite{Hami, Kohl}. 
In the latter case, the current-voltage ($I$-$V$) curve shows step structures at the voltage $V=n(\hbar/2e)f$ with microwave frequency $f$, integer $n$, 
and the ratio of the 
Planck constant and the elementary charge $\hbar/e$. 
This structure called Shapiro step~\cite{Shap} determines the voltage in the order of 10$^{-9}$ accuracy, since the frequency of microwave and the fundamental constants are measured precisely~\cite{Hami, Kohl}.

The Josephson junction composed of SCs and ferromagnet (FM) has been studied by many authors (see Refs.\cite{Buzd,Berg} as review). 
In a {\it ferromagnetic Josephson junction} (FJJ), two superconducting electrodes are separated by a thin ferromagnetic layer.  
By irradiation of microwave, the $I$-$V$ curve of the FJJ shows resonances similar to the Shapiro step~\cite{Hiki08,Houz,Kons,Petk,Volk,Hiki10,Hiki11SST}. It is noted that the resonances appear at the voltages $V=n(\hbar/2e)\Omega$ with ferromagnetic resonance frequency $\Omega$~\cite{Hiki08}. 
Hence, we can see the magnetic response in the ferromagnet by the $I$-$V$ curve of the FJJ~\cite{Hiki11,Hiki13}. 

In this paper, 
we theoretically propose the method to observe the breathing mode of the DW by the FJJ, which has a ferromagnetic layer with the DW between two SCs. 
An equivalent circuit of resistively shunted junction (RSJ) model is used to calculate the $I$-$V$ curve. 
In \S 2, equations of motion for the DW and breathing mode of DW are introduced.
In \S 3, the equivalent circuit of RSJ model is derived by considering the DW in the FJJ, and the \iv~curve is numerically calculated. In addition to the breathing mode, the oscillation of the DW is also examined. 
In \S 4, results are summarized.   

\section{Oscillation and Breathing Mode of Magnetic Domain Wall}
\begin{figure}[ht]
\begin{center}
\includegraphics[width=8cm]{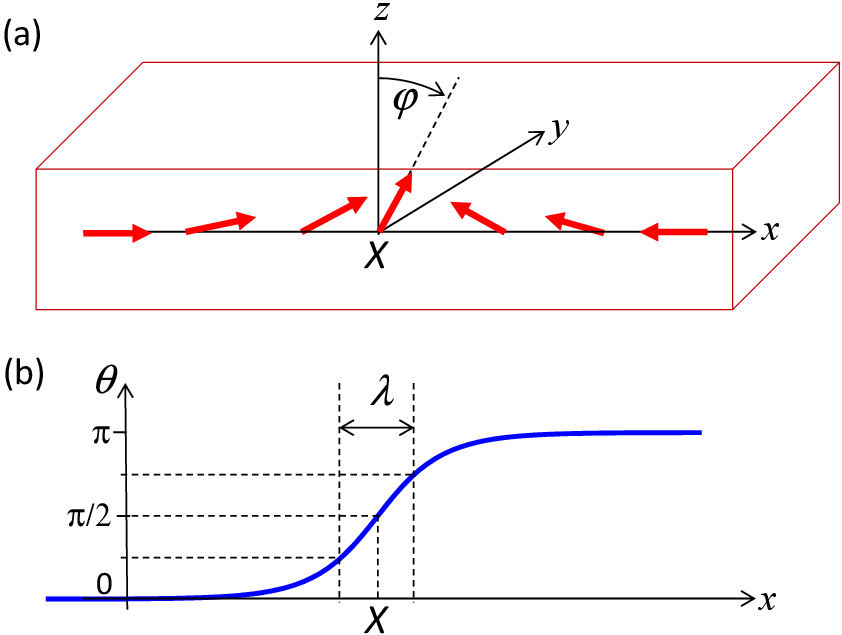}
\caption{ (Color online)
(a) Schematic figure of DW in the FM (rectangle). The red arrows indicate the spatial dependence of the magnetization. A position of the DW is denoted by $X$, and 
$\varphi$ is the tilting angle of the magnetic moment at $X$ from the easy plane ($x$-$z$ plane) to the hard axis ($y$ axis).
(b) Spatial dependence of magnetization determined by an angle 
$\theta=\cos^{-1}\left[\tanh\{(x - X)/\lambda\}\right]$, which is the angle between the magnetic moment and $x$ axis.
}
\label{dw}
\end{center}
\end{figure}

To determine the magnetic domain structure, the geometrical anisotropy is of crucial importance. In this paper, we suppose permalloy for the FM and the junction system with the condition $L_x > L_z$.
In such a nano-wire structure of permalloy, the head-to-head magnetic-domain-structure, which is schematically shown in Fig.~\ref{dw}, is stabilized 
\footnote{
As discussed in the next section, the magnetic flux is restricted in the junction area due to the superconducting diamagnetic current. The hard axis is also brought about by such a junction geometry. 
}.
By external stimuli, the DW shows a dynamics.
In the one-dimensional model, the equation of motion of the DW trapped in pinning potential $V\equiv\beta X^2$ is given by~\cite{Chikazumi,Maekawa,Tata,ieda,Dant,Jung,Mats09,Mats12},
\beqa
&&\frac{{d\varphi }}{{dt}} + \frac{\alpha }{\lambda }\frac{{dX}}{{dt}} = 
- \beta X,\label{eq1}\\
&&\frac{{dX}}{{dt}} - \alpha \lambda \frac{{d\varphi }}{{dt}} = \frac{S}{{2\hbar }}\lambda {K_ \bot }\sin 2\varphi,\label{eq2}\\
&&\alpha \frac{{{\pi ^2}}}{{12}}\frac{{d\lambda }}{{dt}} = \frac{S}{{2\hbar }}\left\{ {\frac{J}{\lambda } - \left( {K + {K_ \bot }{{\sin }^2}\varphi } \right)\lambda } \right\},\label{breath}
\eeqa  
where $X$ is the position of the DW with magnitude of spin $S$. 
The structure of DW is determined by magnetic exchange energy $J$, easy-axis anisotropy $K$ and hard-axis one $K_\perp$. 
Then, the one-dimensional DW with width $\lambda$ is given by 
$\theta  =\cos^{-1}\left[\tanh\{(x - X)/\lambda\}\right]$ as a function of spatial coordinate $x$.
To obtain the breathing mode, the Gilbert damping 
constant $\alpha$ is necessary. 
Here, we note the following two limiting cases~\cite{Tata}; weak pinning case, $V \ll K_\perp$ and strong pinning case, $V \gg K_\perp$. 
In the former case, $\varphi\sim0$ and we obtain the equation 
of motion for $X$ given by, 
\beq
(1 + {\alpha ^2})\frac{{{d^2}X}}{{d{t^2}}} + \alpha \left( {\lambda \beta  + \frac{S}{\hbar }{K_ \bot }} \right)\frac{{dX}}{{dt}} + \lambda \beta \frac{S}{\hbar }{K_ \bot }X = 0,\label{weak}
\eeq
where the DW is associated with an oscillating particle in the pinning potential~\cite{Doring,saitoh}.
This case is relevant to our previous study~\cite{Hiki12}. 
On the other hand, in the latter case, 
the DW is strongly confined around the potential-center (i.e., $X\sim0$)
and the equation of motion for $\varphi$ is obtained as,
\beq
(1 + {\alpha ^2})\frac{{{d^2}\varphi }}{{d{t^2}}} + \alpha \left( {\lambda \beta  + \frac{S}{\hbar }{K_ \bot }\cos 2\varphi } \right)\frac{{d\varphi }}{{dt}} + \lambda \beta \frac{S}{{2\hbar }}{K_ \bot }\sin 2\varphi  = 0. \label{strong} 
\eeq
This results in a rotation or oscillation with respect to the easy-axis.
Note that the oscillation of $\varphi$ induces the breathing mode due to Eq.~(\ref{breath}) (See Refs.~\cite{Dant,Jung,Mats09,Mats12}). 
This naturally derived breathing of the DW
has not been much discussed experimentally,
because of difficulty to build
observation technique.
In the next section, we propose a method sensitive to the breathing mode of the DW.

\section{Equivalent Circuit of RSJ Model with Magnetic Domain Wall}
For a conventional Josephson junction, where two SCs are separated by a thin insulating barrier, 
an equivalent circuit of Josephson junction with the bias current $I$ (RSJ model) is given by, 
\beq
I
=\frac{V}{R} + I_{\rm J}
=\frac{1}{R}\frac{{{\Phi _0}}}{{2\pi }}\frac{{d
\phi }}{{dt}} + {J_{\rm{c}}}\sin 
\phi.\label{RSJ} 
\eeq
Here, the Josephson relation $d\phi/dt = (2e/\hbar)V$ and the flux quantum $\Phi_{0}\equiv h/2e$ are used. 
In the RSJ model, the Josephson junction is associated with a parallel circuit composed of a resistance $R$ and a Josephson current $I_{\rm J}\equiv{J_{\rm{c}}}\sin \phi$ with the Josephson critical current $J_{\rm c}$.  
Because of the gauge invariance in a magnetic field, the phase difference $\phi$ gets the additional term $2\pi\Phi/\Phi_0$ as $\phi\rightarrow\phi+2\pi\Phi/\Phi_0$, where $\Phi$ is the magnetic flux through the junction (See Refs.\cite{Baro,Tink}).

\begin{figure}
\begin{center}
\includegraphics[width=7.5cm]{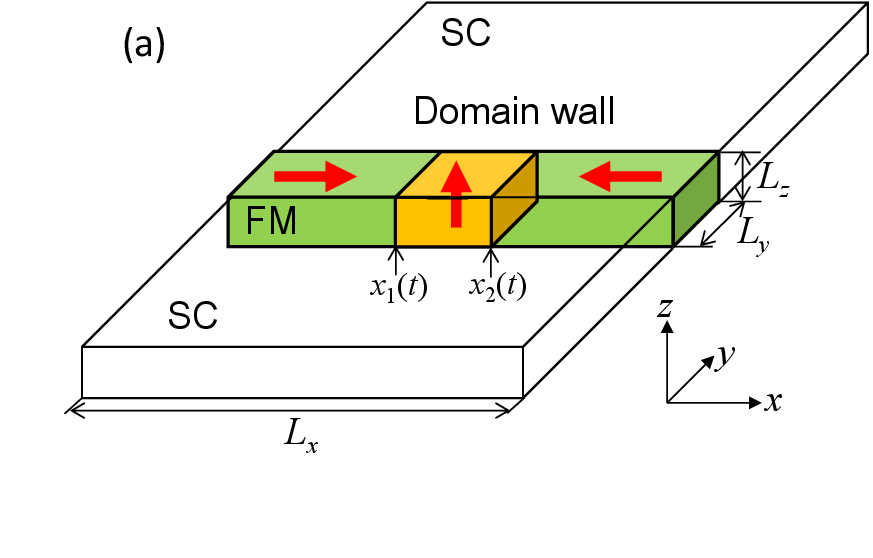}
\includegraphics[width=8cm]{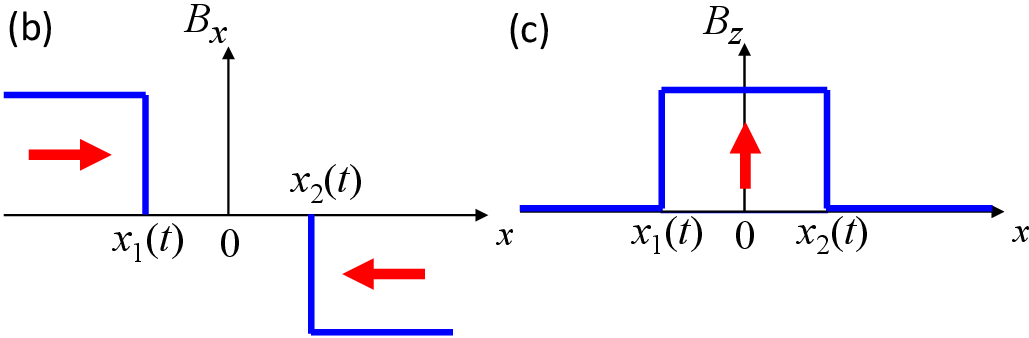}
\caption{ (Color online)
(a) Schematic of an oscillating domain wall in a ferromagnetic layer (FM) that separates two superconductors (SCs). 
Magnetic structure of the FM is shown by the red arrows.
A dc bias current $I$ in the $y$-direction induces a time-dependence of the Josephson phase $\phi$. 
An ac bias current, which induces a breathing mode and/or oscillation of DW is applied to the $x$-direction.
(b) The 
$x$-component and (c) the 
$z$-component of magnetic fields. 
}
\label{fjj}
\end{center}
\end{figure}

We consider the FJJ with DW as shown in Fig.~\ref{fjj}~(a)
\footnote{This simplified structure of DW captures its essential properties and does not change our conclusion.}.
The interface is parallel to the $xz$-plane with width $L_x$ and height $L_z$. A dc bias current is applied in the $y$-direction between the superconducting electrodes separated by depth $L_y$.
The magnetic flux densities $B$ due to the DW is supposed as shown in Figs.~\ref{fjj} (b) and (c), 
where $M_{\rm S}$ is the saturation magnetization and $x_1$ and $x_2$ are both ends of the DW. 
Considering the magnetic flux 
densities induced by the DW  
through the junction, 
$I_{\rm J}$ is given by,  
\beqa
\frac{I_{\rm J}}{J_c}
  &=&
	\left( {\frac{1}{2} + x_1 } \right)\frac{{\sin (\pi {\phi _x})}}{{\pi {\phi _x}}}
	\sin \left(\phi-\pi\phi_z\lambda\right) \nonumber \\
  &+&
	\frac{\sin\left(\pi\phi_z\lambda\right)}{{\pi {\phi _z}}}\sin\left(\phi\right) \nonumber \\
  &+&
	\left( {\frac{1}{2} -x_2} \right)\frac{{\sin (\pi {\phi _x})}}{{\pi {\phi _x}}}
	\sin (\phi  + \pi {\phi _z}\lambda) \label{ij1}\\
  &=&
    \left[ 
		\left(1-\lambda\right) \frac{\sin(\pi\phi_x)}{\pi\phi_x} \cos\left(\pi \phi_z \lambda \right)
       +\frac{\sin \left(\pi \phi_z \lambda \right)}{\pi \phi_z} 
	\right] 
	\sin \left(\phi \right) \nonumber\\
  &-&
	2 x_0 \frac{\sin (\pi \phi_x)}{\pi \phi_x} \sin\left(\pi \phi_z \lambda \right)\cos\left(\phi\right),
\label{ij2} 
\eeqa
with $\phi_{i}\equiv\Phi_{i}/\Phi_{0}$ ($i$ = $x$, $z$), and $\Phi_{x}\equiv L_y L_{z} M_{\rm S}$, $\Phi_{z}\equiv L_{x} L_y M_{\rm S}$~\cite{Hiki12}.
The position of the DW is scaled by $L_x$ and $x_1,x_2 \in [-1/2,1/2]$ 
in units of $L_x$.  
In Eq. (\ref{ij1}), the first and third terms originate from $B_x$ in the regions $x \leq x_1$ and $x_2 \leq x$, respectively,  while the second term is obtained by $B_z$~\cite{Hiki12,Baro,Tink}.  
In Eq. (\ref{ij2}), $(x_1+x_2)/2\equiv x_0$ and $x_2-x_1\equiv\lambda$ are used. 
Note that Eq.~(\ref{ij2}) contains the term proportional to the cosine
function, whereas the critical current in the conventional Josephson
junction is proportional to the sinusoidal function with respect to
the phase difference. The interference of two contributions by the left and the right regions on both sides of the DW is the origin of this term and hence its magnitude is determined by the position of the DW, $x_0$. 
Below, it is assumed that the width of the DW $\lambda$ is smaller than $L_x$. 
In the case 
without DW, i.e., $\lambda=0$, the Josephson critical current depends on the magnetic field in a similar way to the Fraunhofer diffraction pattern as, $J_c \sin(\pi\phi_x)/(\pi\phi_x)$. 
In another case 
with $\lambda=1$, $J_c \sin(\pi\phi_z)/(\pi\phi_z)$ is obtained. 
In both limits, Eq.~(\ref{ij2}) reproduces the Fraunhofer diffraction pattern about the magnetic-field dependence 
on the Josephson critical current. 

We numerically solve the RSJ model with Eq.~(\ref{ij2}) and calculate the $I$-$V$ curve by $V=(\hbar/2e)\langle d
\phi/dt\rangle$~\cite{Baro,Tink}. 
The bracket means the time average, i.e.,  $\langle d
\phi/dt\rangle=(1/T)\int_0^T dt (d
\phi/dt)$. 
We examine the following three cases: (i) The position of DW is
oscillating with a frequency $\nu$,
i.e. $X=x_0+\delta x_0 \sin (\nu t)$
and $\lambda=\lambda_0$~\cite{Hiki12},
namely we assume the oscillation
around $x_0$ with an amplitude $\delta x_0$.
This emulates the weak pinning case,
(see the discussions with Eq.~(\ref{weak})). 
(ii) The width of DW is vibrating
with a frequency $\omega$,
i.e., $X=x_0$ and
$\lambda=
\lambda_0+\delta\lambda\sin(\omega t)$,
where $\lambda_0$ is the mean width of the
DW and $\delta\lambda$ ($<\lambda_0/2$) is the amplitude of the vibration.
This is the breathing mode
and is related to the strong pinning case
(see the discussions with Eq.~(\ref{strong})).
(iii) Both of $X$ and $\lambda$
are time-dependent,
i.e., $X=x_0+\delta x_0 \sin (\nu t)$ and
$\lambda=\lambda_0
+\delta\lambda\sin(\omega t)$\footnote{References~\cite{Dant,Jung,Mats09,Mats12} show the monochromatic oscillation of the DW by applying the electric current. The oscillation will survive even if the Josephson current flows perpendicular to the applied current, since the magnitude of Josephson current is two or three orders 
of magnitude smaller than the applied one~\cite{saitoh}.}.  

\begin{figure}[th]
\begin{center}
\includegraphics[width=7cm]{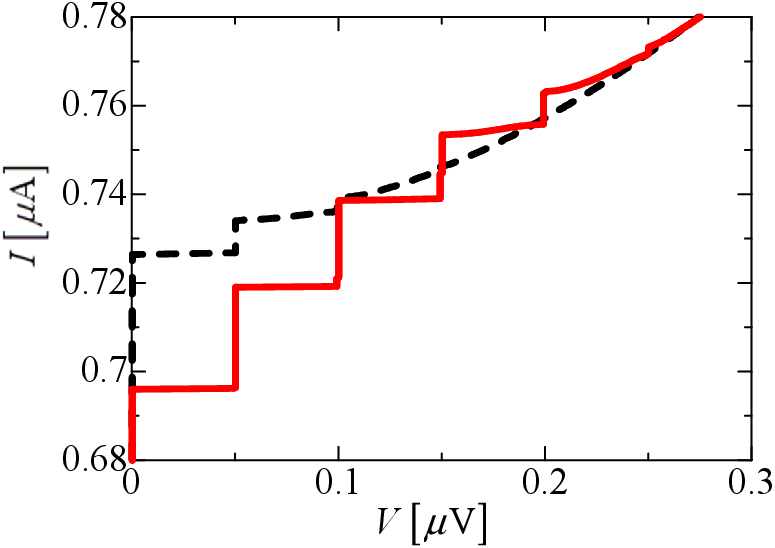}
\caption{ (Color online)
The $I$-$V$ curves of the cases (i) $X=x_0+\delta x_0 \sin (\nu t)$, $\lambda=\lambda_0$ and the case (ii) $X=x_0$, $\lambda=\lambda_0+\delta\lambda\sin(\omega t)$ are shown by the broken (black) and solid lines (red), respectively, for $x_0=0.05$, $\delta x_0=0.1$,  
$\lambda_0 = 0.2$, $\delta\lambda = 0.05$, and $\nu\tau_0$=$\omega\tau_0=0.05$. 
Steps appear at $V=m(\hbar/2e)\nu$ and $V=n(\hbar/2e)\omega$ and  with integer $m$ and $n$. 
Note the relation $V/(J_cR)=n(\omega\tau_0)$ with $(\hbar/2e)(1/J_c R)\equiv\tau_0$. 
}
\label{iv2}
\end{center}
\end{figure}

In Fig. \ref{iv2}, the $I$-$V$ curves of the cases (i) and (ii) are shown by the broken (black) and solid (red) lines, respectively, for $\nu\tau_0=\omega\tau_0=0.05$ with $\tau_0\equiv(\hbar/2e)(1/J_c R)$.
This corresponds to a frequency $\omega=\nu=152$ MHz for the parameters,
$L_{x}$=500 nm, $L_y$=5 nm, $L_{z}$=200 nm, 
$J_{\rm c}$=1 $\mu$A, $R$=1 $\Omega$, $M_{\rm S}$=0.75 T,
$x_0=0.05$, $\delta x_0=0.1$, 
$\lambda_0 = 0.2$, and $\delta\lambda = 0.05$.
We find the step structures at $V=m(\hbar/2e)\nu$ in the broken line (black) and $V=n(\hbar/2e)\omega$ in the solid line (red). 
Note that the step in the solid line (red) is much more clear than that in the broken line (black), even though the magnitude of $\delta x_0=0.1$ is twice as that of $\delta\lambda=0.05$. 
Therefore, our method is
quite sensitive to the breathing mode
of the DW.
Furthermore, our result relates $V$ to $\omega$ ($\nu$) with the fundamental constant $\hbar/e$~\cite{Mohr} \footnote{Accuracy is in the order of $10^{-10}$, e.g., $e$=1.602 176 487(40)~10$^{-19}$C and $\hbar$=1.054 571 628(53)~10$^{-34}$J$\cdot $s.} and integer $n$. 
On the other hand, $V$ is precisely determined by the conventional Josephson junction in the order of $10^{-9}$ accuracy~\cite{Hami, Kohl}. 
The measurement on $V$ determines $\omega$ precisely.  
In addition, the Josephson junction has the highest sensitivity to a magnetic field in the order of femtotesla and the highest speed of switching in the order of tens picosecond~\cite{Baro,Tink}. 
Therefore, our method 
is a powerful method to observe the breathing mode of  
the DW.  

\begin{figure}[th]
\begin{center}
\includegraphics[width=7cm]{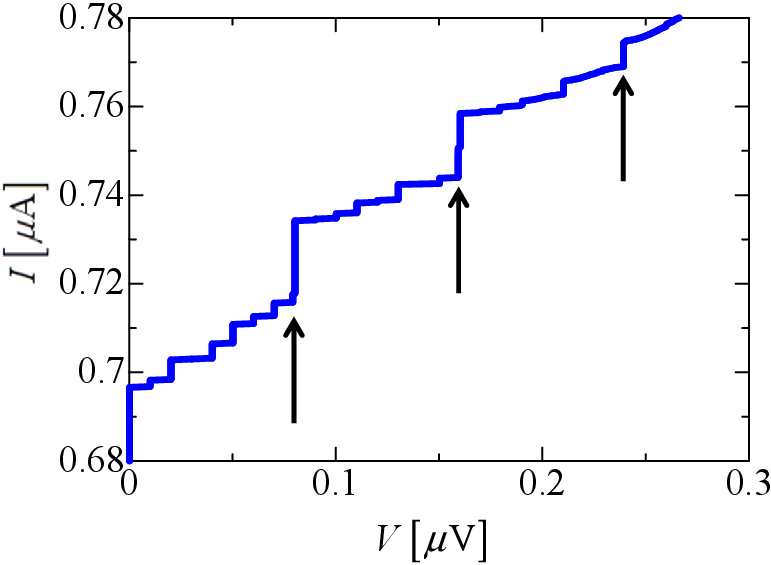}
\caption{ (Color online)
The $I$-$V$ curves of the case (iii) $X=x_0+\delta x_0 \sin (\nu t)$ and $\lambda=\lambda_0+\delta\lambda\sin(\omega t)$, for $x_0=0.05$, $\delta x_0=0.1$,  
$\lambda_0 = 0.2$, $\delta\lambda = 0.05$, $\nu\tau_0=0.05$ and $\omega\tau_0=0.08$. 
The steps occur at
$V=(n \omega + m \nu)(\hbar/2e)$
with integers $n$ and $m$, and the breathing mode is clearly found at the large steps
noted by the arrows, which correspond to $V=n\omega(\hbar/2e)$ with $n$=1, 2, and 3.
}
\label{iv3}
\end{center}
\end{figure}

Figure~\ref{iv3} shows a result of $I$-$V$ curve for the case (iii) with $\nu\tau_0=0.05$ and $\omega\tau_0=0.08$\footnote{Here, we examine more general case with $\nu\neq\omega$, although Eqs~(\ref{eq1}), (\ref{eq2}) and (\ref{breath}) based on the one-dimensional model result in $\nu=\omega$.}. 
Again, we find the step structures
at $V=n(\hbar/2e)\omega$
(due to the breathing of the DW)
and $m(\hbar/2e)\nu$
(due to the oscillation
of the position of the DW).
However, the number of steps is more than expected.
 For example, the step appears at $V$=
0.02, 0.06, and so on.  
These correspond to $V=(\hbar/2e)(n\omega+m\nu)$ with integers $m$ and $n$: 
Here, we consider the following term to elucidate the steps in the $I$-$V$ curve, 
\beq
\sin (\nu t)\sin (\omega t)\cos (\phi ), \label{term}
\eeq
which is involved in Eq.~(\ref{ij2}). 
Because of this time-dependent term, the solution of Eq.~(\ref{RSJ}) 
with Eq.(\ref{ij2}) contains the component given by, 
\beq
\phi  = (\frac{2e}{\hbar})Vt + A\sin ({\omega}t) + B\sin ({\nu}t), \label{sol}
\eeq
where $A$ and $B$ are constants. 
Considering Eq.~(\ref{sol}), we find that Eq.~(\ref{term}) is composed of the following terms,     
\beq
e^{ \pm i\phi  \pm i(\omega  \pm \nu )t} 
=\sum_{n,m=-\infty}^\infty  {J_n\left( A \right)J_m\left( B \right)e^{i(2e/\hbar)Vt - in\omega t - im\nu t}}, \label{component}  
\eeq
where $J_n(A)$ and $J_m(B)$ are 
the Bessel functions of the first kind.  
The step structures are associated with the dc component of $I_J$, which is given by taking the time average
of Eq.~(\ref{ij2}). 
Then, the constant term in Eq.~(\ref{component}), i.e., time-independent term, gives the step structure in the \iv~curve. 
Therefore, we find the condition
$(2e/\hbar)V-n\omega-m\nu=0$
for the step structures. 
In particular, the steps appear at $V$=
0.02, and 0.06 for $(n,m)$=$(-1,2)$, and $(2,-1)$, respectively.
Although the step structures due to
the oscillation and the breathing mode
of the DW are mixed up in the \iv curve, the breathing mode can be identified by the large steps even in the case (iii) as noted by the arrows in Fig.~\ref{iv3}.  

\section{Summary and Discussions}
We have theoretically proposed how to observe the breathing mode 
of DW by the Josephson junction having a ferromagnetic layer with the magnetic domain wall between superconducting electrodes. 
The current-voltage ($I$-$V$) curve is calculated by an equivalent circuit of the resistively shunted junction model. 
The breathing mode is identified by stepwise structures in the $I$-$V$ curve, which appear at the voltages $V=n(\hbar/2e)\omega$ with the fundamental constant $\hbar/e$, integer number $n$, and the frequency of the breathing mode $\omega$. 
This is the most feasible  
method to observe the breathing mode of the DW due to the sensitivity of the Josephson junction. 

In principle, magnetic resonance methods are also possible to measure the breathing mode. 
Owing to their low sensitivity, however, the resonance measurements need a sample containing a  number of DWs, and furthermore those DWs need to vibrate collectively.  
Although such a measurement is realized in the Skyrmion lattice~\cite{Onos,Moch}, in which the rotational and the breathing modes are observed, it has not been done on the breathing mode of the DW so far. 
On the other hand, our method 
is accessible to the breathing mode of a single DW due to the sensitivity of the Josephson junction~\cite{Baro,Tink}. 

The authors would like to thank J. Ieda for useful discussions and comments. 
This work was partly supported by Grant-in-Aid for Scientific Research from MEXT (Grant No.24540387, No.24360036, No.23340093, and No.25287094) and by the inter-university cooperative research program of IMR, Tohoku University.
Numerical computation in this work was carried out on the supercomputers at JAEA.


\begin{thebibliography}{}
\bibitem{Chikazumi}
S. Chikazumi,
\textit{Physics of Ferromagnetism} (Oxford University Press, New York, 1997).

\bibitem{Maekawa}
S. Maekawa, 
\textit{Concepts in Spin Electronics} (Oxford University Press, Oxford, 2006). 

\bibitem{Doring}
W. D\"oring, 
Z. Naturforsch {\bf 3A} 373 (1948).

\bibitem{saitoh}
E. Saitoh, H. Miyajima, T. Yamaoka, G. Tatara, 
Nature {\bf 432}, 203 (2004).

\bibitem{Yamano}
M. Yamanouchi, D. Chiba, F. Matsukura, and H. Ohno, 
Nature {\bf 428}, 539 (2004).

\bibitem{Yamagu}
A. Yamaguchi, T. Ono, S. Nasu, K. Miyake, K. Mibu, and T. Shinjo, 
Phys. Rev. Lett. {\bf 92}, 077205 (2004).

\bibitem{Thomas}
L. Thomas, M. Hayashi, X. Jiang, R. Moriya, C. Rettner and S. Parkin,
Nature {\bf 443}, 197 (2006).

\bibitem{hayashi}
M. Hayashi, L. Thomas, C. Rettner, R. Moriya and S. S. P. Parkin, 
Nature Physics {\bf 3}, 21 (2007).

\bibitem{boone}
C. T. Boone, J. A. Katine, M. Carey, J. R. Childress, X. Cheng, and I. N. Krivorotov, 
Phys. Rev. Lett. {\bf 104}, 097203 (2010). 

\bibitem{bisig}
A. Bisig, L. Heyne, O. Boulle, M. Kl\"aui, 
Appl. Phys. Lett. {\bf 95}, 162504 (2009).

\bibitem{Thia}
A. Thiaville, J. M. Garc\'ia, and J. Miltat, 
J. Magn. Magn. Mater. {\bf 242-245}, 1061 (2002).

\bibitem{he07}
J. He and S. Zhang,
Appl. Phys. Lett. {\bf 90}, 142508 (2007).

\bibitem{Tata}
G. Tatara, H. Kohno, and J. Shibara,
Phys. Rep. {\bf 468}, 213 (2008).

\bibitem{ieda}
J. Ieda, H. Sugishita, and S. Maekawa, 
J. Magn. Magn. Mat. {\bf 322}, 1363 (2010).

\bibitem{Hiki12}
S. Hikino, M. Mori, W. Koshibae, and S. Maekawa, 
Appl. Phys. Lett. {\bf 100}, 152402 (2012).

\bibitem{Dant}
A. L. Dantas, M. S. Vasconcelos,
and A.S. Carri\c{c}o,
J. Magn. Magn. Mater. {\bf 226-230}, 1604 (2001).

\bibitem{Jung}
S. W. Jung and H. W. Lee,
J. Magnetics {\bf 12}, 1 (2007).  

\bibitem{Mats09}
K. Matsushita, J. Sato, and H. Imamura, 
J. Phys. Soc. Jpn. {\bf 78}, 093801 (2009).

\bibitem{Mats12}
K. Matsushita, M. Sasaki, J. Sato, and H. Imamura, 
J. Phys. Soc. Jpn. {\bf 81}, 043801 (2012).

\bibitem{Baro}
A. Barone and G. Patern\'o, 
\textit{Physics and Applications of the Josephson Effect} (Wiley, New York, 1982).

\bibitem{Tink}
M. Tinkham, 
\textit{Introduction to superconductivity (Second Edition)} (Dover, New York, 2004).

\bibitem{Hami}
C. A. Hamilton, 
Rev. Sci. Instrum. {\bf  71}, 3611 (2000). 

\bibitem{Kohl}
J. Kohlmann, R. Behr, and T. Funck, 
Meas. Sci. Technol. {\bf 14}, 1216 (2003).

\bibitem{Shap}
S. Shapiro, 
Phys. Rev. Lett. {\bf 11}, 80 (1963).

\bibitem{Buzd} 
A. I. Buzdin, 
Rev. Mod. Phys. {\bf 77}, 935 (2005). 

\bibitem{Berg}
F. S. Bergeret, A. F. Volkov, and K. B. Efetov, 
Rev. Mod. Phys. {\bf 77}, 1321 (2005).

\bibitem{Hiki08}
S. Hikino, M. Mori, S. Takahashi, and S. Maekawa,
J. Phys. Soc. Jpn. {\bf 77} 053707 (2008).

\bibitem{Houz}
M. Houzet, 
Phys. Rev. Lett. {\bf 101}, 057009 (2008).

\bibitem{Kons}
F. Konschelle and A. Buzdin, 
Phys. Rev. Lett. {\bf 102}, 017001 (2009).

\bibitem{Petk}
I. Petkovi\'c, M. Aprili, S. E. Barnes, F. Beuneu, and S. Maekawa, 
Phys. Rev. B {\bf 80}, 220502 (2009). 

\bibitem{Volk}
A. F. Volkov and K. B. Efetov, 
Phys. Rev. Lett. {\bf 103}, 037003 (2009). 

\bibitem{Hiki10}
S. Hikino, M. Mori, S. Takahashi, and S. Maekawa, 
Physica C {\bf 470}, S819 (2010).

\bibitem{Hiki11SST}
S. Hikino, M. Mori, S. Takahashi, and S. Maekawa, 
Supercond. Sci. Technol. {\bf 24}, 024008 (2011).

\bibitem{Hiki11}
S. Hikino, M. Mori, S. Takahashi, and S. Maekawa, 
J. Phys. Soc. Jpn. {\bf 80}, 074707 (2011).

\bibitem{Hiki13}
S. Hikino, M. Mori, and S. Maekawa, 
arXive:1306.3652


\bibitem{Vers}
J. J. Versluijs, M. A. Bari, and J. M. D. Coey,
Phys. Rev. Lett. {\bf 87}, 026601 (2001).

\bibitem{Barn}
S. E. Barnes, J. Ieda, and S. Maekawa, 
Appl. Phys. Lett. {\bf 89}, 122507 (2006). 

\bibitem{Mohr}
P. J. Mohr, B. N. Taylor, and D. B. Newell, 
Rev. Mod. Phys. {\bf 80}, 633 (2008). 

\bibitem{Onos}
Y. Onose, Y. Okamura, S. Seki, S. Ishiwata, and Y. Tokura,
Phys. Rev. Lett. {\bf 109}, 037603 (2012).

\bibitem{Moch}
M. Mochizuki,
Phys. Rev. Lett. {\bf 108}, 017601 (2012).
\end{thebibliography}
\end{document}